\documentclass[11pt,twoside]{article}

\usepackage{asp2006}
\usepackage{lscape}
\usepackage{graphicx}

\begin{document}
\title{Ellipsoidal effect in the symbiotic star YY Her}

\author{M.~Wi\c{e}cek, M.~Miko\l{}ajewski, T.~Tomov, E.~\'Swierczy\'nski, C.~Ga\l{}an, P.~Wychudzki}

\affil{
\footnotesize
Centre for Astronomy of Nicolaus Copernicus University, Toru\'n, Poland}

\begin{abstract} 
A new estimation of the orbital period of YY Her on the base of our and published observations is presented.
Phased light curves in $(RI)_{\rm C}$ bands show evidently ellipsoidal effect connected with the tidal distortion of the giant surface.
\end{abstract}

\section*{Photometric data}
We obtained  $\rm {UBV(RI)_{\rm C}}$ photometry of YY~Her, using a 60\,cm Cassegrain telescope at Piwnice Observatory near
Toru\'n (Poland), equipped with  an EMI~9558B photomultiplier (1991--1999), a RCA~C31034 photomultiplier (2001--2004) and a SBIG STL 1001 CCD camera (2005--2008).
Additionally, we used data published by Hric et al. (2006), Tatarnikova et al. (2001), Miko\l{}ajewska et al. (2002) and data from ASAS (Pojmanski 2002).

To eliminate the systematic shifts between the different photometric systems, some data sets were corrected as follows: Tatarnikova et al. ($-0\fm52$, $-0\fm16$, $+0\fm3$, $+0\fm5$) 
and Miko\l{}ajewska et al.($-0\fm22$, $-0\fm1$, $+0\fm27$, $+0\fm53$) in $B, V, R_{\rm C}, I_{\rm C}$ respectively.
\section*{Analysis of the light curves and the orbital period estimation}
The multicolor photometric observations of YY~Her are presented in Figure~1 (left panel).
Fast Fourier Transform was used to search for the orbital period. The corresponding periodograms for all pass-bands are shown in Figure~1 (middle panel). 
Two peaks at about $\sim 575^{\rm d}$ and $\sim 284^{\rm d}$ dominate in these periodograms. The first one is not visible in the $I_{\rm C}$ filter and the second one
 is absent in the $U$ filter.
Assuming that the highest peak corresponds to the orbital period, to estimate its mean value ($575\fd75$) we used the peaks in $U, B, V, R_{\rm C}$ filters. 

Using this mean value and measuring the moments of the primary minima from the $V$ light curve, we constructed the O-C diagram which
 allowed us to introduce a correction of $0\fd23$ for  $\Delta T_{\rm 0}$ and $-1\fd17$ for $P_{\rm orb}$. 
Finally, we adopted the ephemeris $JD(I)_{\rm min} = 2450702\fd75 + 574\fd58 \times E $.

The photometric data in all filters, phased with our ephemeris are shown in Figure~1 (right panel). The $U$ light curve shows a pure sine wave shape with a large
 amplitude $> 0\fm8$ and probably reflects the eclipse of the ionized HII zone by red giant and neutral HI region in the system. The second minimum appears in the $B$ 
light curve at orbital phase $0.5$ and is very well seen in the $V, (RI)_{\rm C}$ ones. We estimated its mean value $284\fd47$ 
using the peaks in the $V, (RI)_{\rm C}$ periodograms. We suggest that this second minimum is caused by ellipsoidal changes of the red giant.

To measure the moments of the secondary minima, we used the $I_{\rm C}$ light curve in which these minima are best visible.
 Assuming $T_{\rm 0}=2450693\fd2$ and using the secondary period mean value ($284\fd47$) we estimated the corrections 
$\Delta T_{\rm 0}=+6\fd4$ and $\Delta P_{\rm orb}=-0\fd32$. Our final ephemeries for the second minimum is $JD(II)_{\rm min} = 2450699\fd6 + 284\fd15 \times E$.
Double secondary period ($568\fd3$) is $\sim 6.3$ days shorter than the orbital period. This is a significant difference which will be analysing in the future.
\begin{figure}[!htp]
  \centering  
 \includegraphics[width=1.0\textwidth, angle=0]{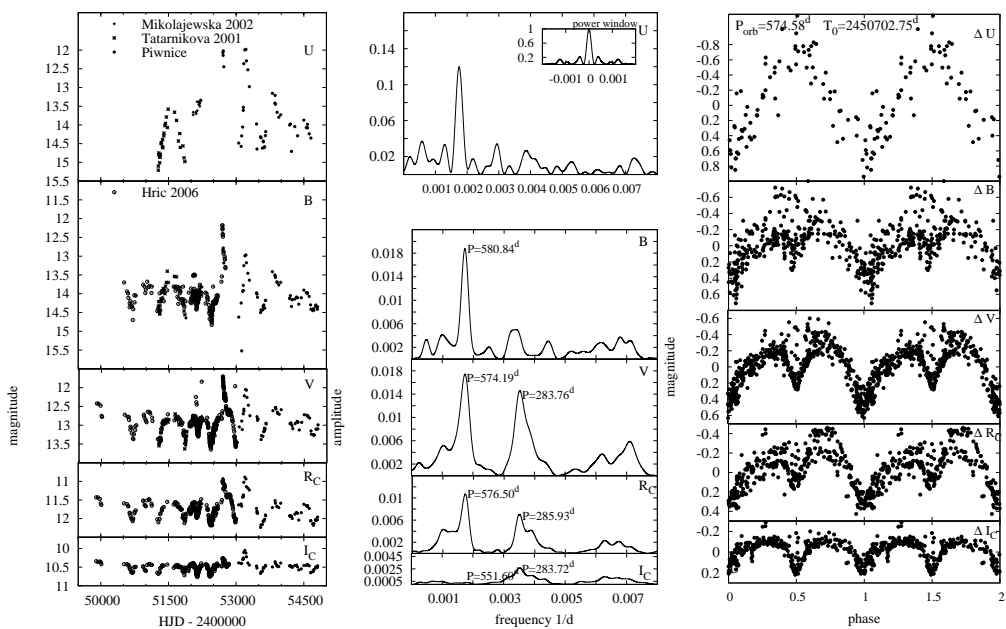}
 \caption{Left panel: Multicolour light curves of YY~Her,
Middle panel: Power spectrum for $UBV(RI)_{\rm C}$ filters data of YY~Her,
Right panel: The residual $UBV(RI)_{\rm C}$ light curves of YY~Her phased
with $574\fd58$ period.}
  \label{rys1}
\end{figure}
\acknowledgements This work is supported by the Polish MNiSW Grant N203~018~32/2338 and European ZPORR project in Kujawsko-Pomorskie Region "exhibitions for PhD students".


\begin{thebibliography}{}
\bibitem[Hric et al. (2006)]{hric_2006} Hric L., G\'{a}lis R., Niarchos, P., Dobrotka, A., \v{S}imon, V., \v{S}melcer, L., Veli\u{c}, Z.,H\'{a}jek, P., Gazeas, K., Sobotka, P., Koss, K., 2006, \textit{CAOSP} \textbf{36},26 
\bibitem[Miko\l{}ajewska et al. (2002)]{miko_2002} Miko\l{}ajewska, J., Kolotilov, E.A., Shugarov, S.Yu., Yudin, B.F, 2002, \textit{A\&A} \textbf{392}, 197
\bibitem[Pojma\'nski (2002)]{pojma_2002} Pojmanski, G., 2002, \textit{Acta Astronomica}, \textbf{52}, 397
\bibitem[Tatarnikowa et al. (2001)]{tatar_2001} Tatarnikova, A.A., Esipov, V.F., Kolotilov, E.A., Miko\l{}ajewska, J., Munari, U., Shugarov, S.Yu., 2001, \textit{Astronomy Letters} \textbf{27}, 11 
\end{thebibliography}
\end{document}